Crater-ray formation through mutual collisions of hypervelocity-impact induced ejecta particles

Toshihiko Kadono[a,*], Ayako I. Suzuki[b], Rintaro Matsumura[a], Junta Naka[a], Ryo Suetsugu[a], Kosuke Kurosawa[c], and Sunao Hasegawa[b]

[a] *Department of Basic Sciences, University of Occupational and Environmental Health, Kitakyusyu 807-8555, Japan*

[b] *Institute of Space and Astronautical Science, Japan Aerospace Exploration Agency, Kanagawa 252-5210, Japan*

[c] *Planetary Exploration Research Center, Chiba Institute of Technology, Chiba 275-0016, Japan*

* Corresponding author.

*E-mail address:* kadono@med.uoeh-u.ac.jp (T. Kadono)

**ABSTRACT**

We investigate the patterns observed in ejecta curtain induced by hypervelocity impact (2−6 km/s) with a variety of the size and shape of target particles. We characterize the patterns by an angle, defined as the ratio of the characteristic length of the pattern obtained by Fourier transformation to the distance from the impact point. This angle is found to be almost the same as that obtained by the reanalysis of the patterns in the previous study at lower impact velocities (Kadono et al., 2015, Icarus 250, 215-221), which are consistent with lunar crater-ray systems. Assuming that the pattern is formed by mutual collision of particles with fluctuation velocity in excavation flow, we evaluate an angle at which the pattern growth stops and show that this angle is the same in the order of magnitude as the ratio of the fluctuation velocity and the radial velocity. This relation is confirmed in the results of experiments and numerical simulations. Finally, we discuss the dependence of the patterns on impact conditions. The experiments show no dependence of the angle on impact velocity. This indicates that the ratio between the fluctuation and radial velocity components in excavation flow does not depend on impact velocity. Moreover, the independences on particle size and particle shape suggest that the angle characterizing the structure of the patterns does not depend on cohesive force. Since cohesive forces should be related with elastic properties of particles, the structure does not depend on elastic properties, though inelastic collisions are important for the persistence and contrast of the patterns.

**1.Introduction**

Crater rays are typically observed around craters on solid surfaces of celestial bodies such as the moon (see e.g., Kaguya website: http://www.kaguya.jaxa.jp/gallery/index_e.html). The nature of the rays has been studied for a long time, and the origin of the brightness of the rays is fairly understood (e.g., Hawke et al., 2004). Another remarkable feature of crater rays, the spatial non-uniformity, is recently paid attention as a geologic process such that the spatially heterogeneous rays determine the rate of degradation of small craters (Minton et al., 2019). However, the mechanism of non-uniformity of crater-rays remains an unsolved problem; there are only a few studies investigating the mechanism for the generation of non-uniformities of the rays such as high-velocity detonation products in explosion cratering experiments analogous to the impact-induced vapor (Andrews, 1977), interaction of shock waves with old craters (Shuvalov, 2012), pattern formation through mutual collisions in granular media (Kadono et al., 2015), and, effect of the topography around the impact point (Sabuwala et al., 2018).

Impact on planetary and lunar surfaces is expected to occur at hypervelocity (> ~km/s). However, in our previous experiments the results at the impact velocities lower than ~100 m/s were shown (Kadono et al. 2015). The impact velocity was also low in the experiments by Sabuwala et al. (2018) (< ~10 m/s; projectiles were freely falling). Therefore, there has been no experimental studies on the non-uniformity of crater-rays at hypervelocity expected in natural impacts.

In this paper, we investigate the nonuniform structures observed in ejecta curtain induced by hypervelocity impact (2−6 km/s). First, we describe the results of hypervelocity impact experiments with granular targets with a variety of the particle size and shape, in particular, the spatial distributions of granular materials during their flight after impact by taking consecutive images of the ejecta curtain using a high-speed camera. Then, we analyze the patterns observed in ejecta curtain using Fourier transformation. As one possible process to form such non-uniform distribution in the ejecta curtain, mutual collision of particles was suggested (e.g., Lohse et al., 2004; Kadono et al., 2015), based on the result that inelastic mutual collision of particles plays an important role to spontaneously form inhomogeneous clustering state (e.g., Luding and Herrmann, 1999; Goldhirsch, 2003). Therefore, assuming that collision process underlies the pattern formation, we derive the relation between the patterns and the fluctuation velocity with the support of numerical simulations and finally discuss the dependence of the patterns on impact conditions.

## 2.Experiment

We carried out impact experiments using a two-stage hydrogen-gas gun at Institute of Space and Astronautical Science, JAXA. Spherical polycarbonate projectiles with a diameter of 4.8 mm (0.068 g in mass) were accelerated, which impacted perpendicularly onto the surface of granular targets. Five shots were made: a spherical glass bead target with a diameter of 0.04−0.05 mm at an impact velocity of 2.3 km/s (denoted as GB50), spherical glass beads with a diameter of 0.09−0.11 mm at impact velocities of 2.2 km/s (GB100_2) and 5.9 km/s (GB100_6), spherical glass beads with a diameter of 0.25−0.36 mm at impact velocities of 2.3 km/s (GB300), and irregularly shaped silica ($SiO_2$) sands with a diameter of 50−100 μm at an impact velocity of 2.3 km/s (SS70). The targets were poured into a bowl, which had a radius of 15 cm and depth of 10 cm with a flat bottom, and set in a vacuum chamber. The ambient pressure was less than 2.5 Pa. The motion of the ejecta curtain was observed using a high-speed video camera (HPV-X, Shimadzu Co.

Ltd) with a rate of 2 ms per frame. After the shot, we measured the rim-to-rim diameter of final craters. A schematic diagram of the experimental configuration is shown in Fig.

1.

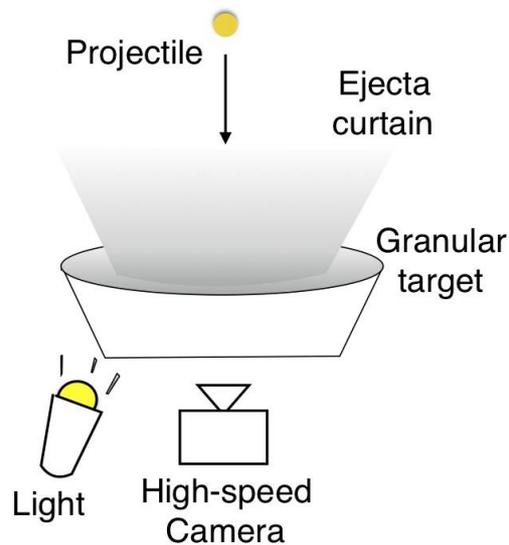

**Fig. 1.** Schematic diagram of the experimental configuration. A polycarbonate projectile

impacted perpendicularly onto the target surface from the top in a vacuum chamber. A high-speed camera and a light were set outside of the chamber.

### 3.Results

   Figure 2 shows consecutive images of GB100_2 (the movie for each shot is available in the Supplemental Materials (SM)). A projectile coming from the top impacted perpendicularly onto the surface of the granular target and the spatial distribution of the ejected particles from each shot showed a nonuniform pattern.

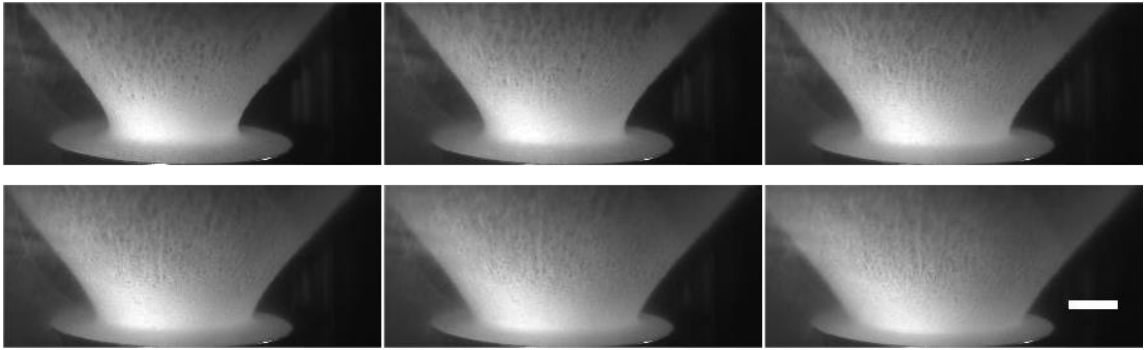

**Fig. 2.** Consecutive images of ejecta curtain for a glass beads target GB100_2. A polycarbonate projectile impacted perpendicularly onto the target surface from the top. The upper-left panel shows the ejecta curtain at 34 ms after the impact and the time proceeds from the upper-left to -right and then lower-left to -right panels with an interval of 10 ms. A white horizontal line in lower-right panel indicates a spatial scale of 50 mm.

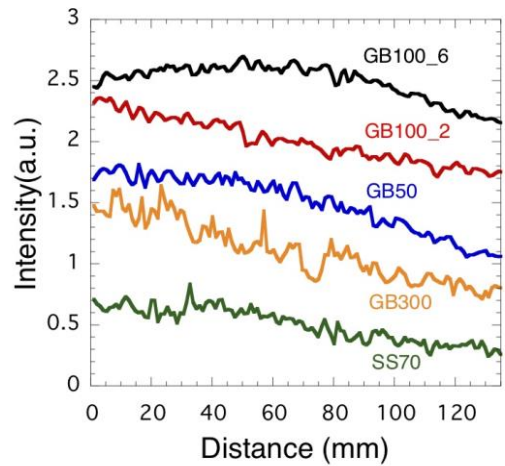

**Fig. 3.** Intensity distributions at 44 ms after the impact of GB100_6 (black line), GB100_2 (red), GB50 (blue), GB300 (orange) and SS70 (green). The curves are offset from each other to avoid crossovers.

We measured the distribution of intensities along a horizontal line on the ejecta curtain (the measured lines are shown in Figs. S1a−e in the SM) with a length of 128 pixels (163 mm). Note that the length between two points along the horizontal arcs in the ejecta curtain is different from that projected on a plane (as images taken by the camera). Hence, we corrected the intensity data taken by the camera in order to obtain intensities at points evenly distributed along the horizontal arcs in the ejecta curtain (see the SM). The corrected distributions are shown in Fig. 3, where the horizontal axis denotes the spatial length along the horizontal arc in the ejecta curtain. The distributions have an oscillatory behavior; hence, we analyzed the distributions using Fourier transformation to investigate their periodic structures. The spectra were plotted for every 10 ms starting from 30 ms after the impact; each spectrum is an average of the spectra for five consecutive frames (e.g., 30, 32, 34, 36, and 38 ms) (the spectra for all of the shots are shown in Fig. S3 in the SM). One example of the obtained power spectra is shown in Fig. 4 for GB100_2, where the vertical axis denotes the square of the amplitude $F$ and the horizontal axis denotes the frequency $k$ in units of $mm^{-1}$. Note that only the data for $k$ between 0.05 and 0.3 $mm^{-1}$ were plotted. The spectra at large wavenumbers of 49−64 ($k > \sim 0.30$ $mm^{-1}$) were considered as noise and the counts averaged over this range were subtracted from the data. In addition, the spectra at small wavenumbers of 1−9 ($k < \sim 0.05$ $mm^{-1}$), corresponding to $\sim 1/10$ of the length of the line that we measured, were not considered because they may include the influence of the length of the line that we measured. The power ($F^2$) in the spectra decreases with $k$ and does not show clear peaks even though the scatter is large, which is consistent with the findings of a previous study (Kadono et al., 2015).

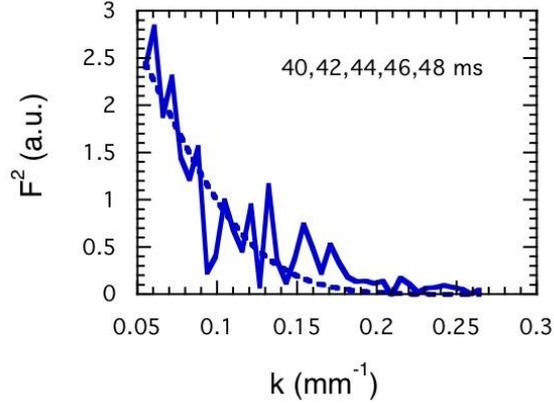

**Fig. 4.** A power spectrum for GB100_2 as a function of frequency $k$ (mm$^{-1}$). This spectrum represents an average of the spectra for five consecutive frames during 40−48 ms (i.e., 40, 42, 44, 46, and 48 ms) after the impact. The characteristic length $\lambda_c$ is obtained by fitting a Gaussian function to the spectra. The fitting curve is also shown in the figure as a dotted curve.

In the investigation of cluster formation process in a freely evolving gas and the density correlation function, it is found that the system evolution is described by a single length scale and that the asymptotic dependence of this length on time is consistent with diffusive growth (Das and Puri, 2003). Based on this result, we suppose that diffusion is fundamental for the pattern formation in the ejecta curtain and fit a Gaussian function $AExp[−Bk^2]$ to the power spectra in Fig. 4, where $A$ and $B$ are the fitting parameters. Parameter $B$ corresponds to $\lambda_c^2$, where $\lambda_c$ is the characteristic length of the patterns of ejecta curtain. Figure 4 shows the fitting curve alongside the data, which demonstrates the suitability of the Gaussian function for the experimental data. We consider an "angle" $\theta_c$ at which the impact point takes a view of $\lambda_c$, defined as $\lambda_c/r$, where $r$ is the distance from the impact point to a point on the horizontal arc that we measured on the ejecta curtain. Using an average value of $r$ during an interval of 10 ms, $\theta_c$ is obtained. Figure 5 shows $\theta_c$ as a function of time normalized by the characteristic crater formation time $\tau = (D_c/g)^{1/2}$, where $D_c$ is the rim-to-rim diameter of the final crater (~13.5−26.0 cm in our experiments) and $g$ is the acceleration due to gravity (Melosh, 1989). It is noted that $\tau$ becomes ~100 ms in our case, and hence, the results shown correspond to the intermediary and later stages of crater formation. We also show the result of a previous study using 100-μm glass beads at a lower impact velocity of ~100 m/s (denoted as GB100_100), obtained by the reanalysis of the images during 30−90 ms after the impact shown in Fig. 3 of Kadono et al. (2015) with the same procedure as that in this paper (i.e., fitting a Gaussian function to the noise-subtracted power spectra); we exclude the data from the images after 90 ms because they are slightly out of the focus of the imaging optics. Results show that $\theta_c$ remains almost constant ~0.06 rad across time after the impact (though the result with GB300 largely scatters at the later stages (> ~0.5) due to the sparseness of ejecta distribution and the low contrast of images, the average during the stages is similar to that at the earlier stages). A constant angle implies that the pattern expands geometrically and does not evolve; the formation process in the ejecta curtain has finished at the observation time, which is consistent with a previous study (Kadono et al., 2015). The fact that no difference in $\theta_c$ is found among the results of GB100_100, GB100_2, and GB100_6 indicates that no dependence of $\theta_c$ is expected on impact velocity. Moreover, no difference among the results of GB50, GB100_2, and GB300 suggests that no dependence of $\theta_c$ is expected on particle sizes. On the other hand, the results show that $\theta_c$ for silica sands appears to be slightly lower than that for glass beads at the later stages, but the difference of the data seems to be within their scatters. In fact, the averages of SS70 and GB50 during the common time range shown in Fig. 5 ($0.3 < \tau < 0.8$) are $0.051\pm0.009$ and $0.065\pm0.007$, respectively, suggesting that there is no significant difference (the difference in the previous results at lower impact velocities is also small within the errors (Kadono et al., 2015)).

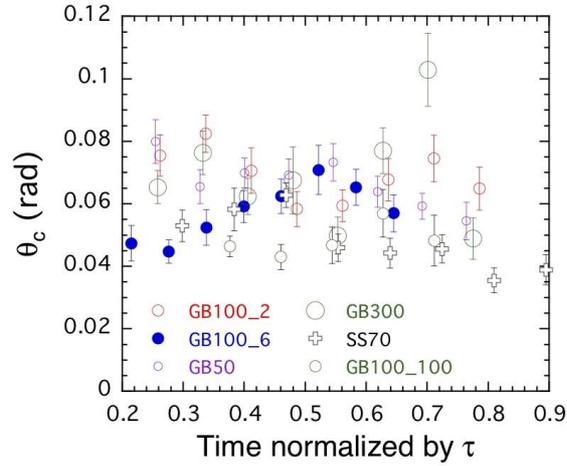

**Fig. 5.** Angle $\theta_c$ (in units of radian), defined as the ratio of the characteristic length $\lambda_c$ to the distance from the impact point, is plotted as a function of time after impact normalized by the characteristic crater formation time $\tau = (D_c/g)^{1/2}$. Since $\lambda_c$ is obtained from the averaged power spectra for an interval of 10 ms, $\theta_c$ is plotted at the median of each range. The errors in $\theta_c$ are mainly from those in $\lambda_c$ caused by the fitting.

## 4. Discussion

### 4.1. $\theta_c$ and inelastic collision process

Here, we consider target granular materials in excavation flow and look at a hypothetical horizontal circle getting on the excavation flow centered at impact point with a radius of $r(t)$, which increases with time $t$ with a velocity of $v_r(t)$. When the particles on the circle having a fluctuation velocity perpendicular to the radial direction (i.e., along the circle) $\Delta v(t)$, cluster formation occurs through inelastic collisions on the circle. Length scale between clusters on the circle $\lambda(t)$ increases with time as clustering proceeds. The experimental result that $\theta_c$ is constant in time implies that cluster growth stops at a finite time. To represent this situation, we suppose the circumference length of the circle geometrically becomes so large in time that particle collisions along the circle cease. The circumference length of the circle $L(t) = 2\pi r(t)$ increases with time due to the geometrical expansion as $dL(t)/dt = 2\pi v_r(t)$ and the relative velocity $v_{rel}(t)$ between two points separated by $\lambda(t)$ along the circle is $v_{rel}(t) = \{\lambda(t)/L(t)\}(2\pi v_r(t)) = \{\lambda(t)/r(t)\}v_r(t)$. Supposing that collision due to diffusion stops when $\Delta v(t) \sim v_{rel}(t)$, we obtain the characteristic angle $\theta_c$, at which particle accumulation through collisions stops, as $\theta_c = \lambda(t^*)/r(t^*) \sim \Delta v(t^*)/v_r(t^*)$, where $t^*$ is the time at which the collision stops. This indicates that $\theta_c$ is the same order of magnitude as the ratio of the fluctuation velocity to the radial velocity of excavation flow.

### 4.2. $\theta_c$ versus the ratio of fluctuation velocity to radial velocity

#### 4.2.1. Experiment

In GB300, spatially scattered particles were observed to be apart about an order of $\delta$ ~several mm from the ejecta curtain at $r$ ~10–20 cm in distance from the impact point (Fig. 6). Moreover, in experiments using slightly larger glass spheres (~500 μm), spatially scattered particles were also observed to be $\delta \sim$ a few mm and $r \sim$ 10–20 cm (Tsujido et al., 2015). This scattering was caused by velocity fluctuation in the direction of the zenith angle $v_\varphi$. The ratio of the fluctuation velocity $v_\varphi$ to radial velocity $v_r$ of particles is evaluated to be $\delta/r \sim$ 0.01–0.1, which is consistent with a value of $\sim$0.01–0.1 obtained from the previous experiments with freely falling granular streams (Amarouchene et al., 2008; Royer et al., 2009). Assuming $v_\varphi$ and $v_r$ to be $\Delta v(t^*)$ and $v_r(t^*)$, respectively, the ratio $\Delta v(t^*)/v_r(t^*)$ is $\sim$0.01–0.1 and actually the same in the order of magnitude as $\theta_c$ obtained in our experiments as shown in Fig. 5.

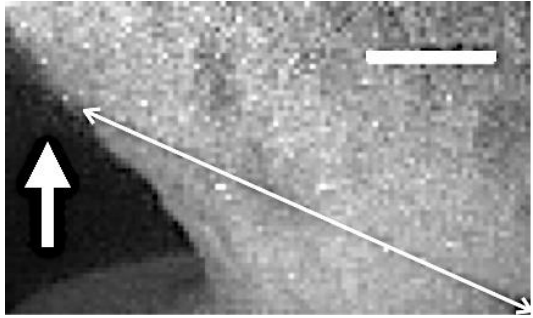

**Fig. 6.** Ejecta curtain in the target of GB300 at 62 ms after the impact. Some scattered particles can be recognized (indicated by a bold arrow). Horizontal bar indicates 30 mm and a narrow line shows the distance from the impact point to the ejecta curtain.

### 4.2.2. Numerical simulation

To confirm alternatively the relation between $\theta_c$ and $\Delta v(t^*)/v_r(t^*)$, we numerically investigated the pattern in impact-induced ejecta curtain using an open-source discrete-element-method simulator, LIGGGHTS (Kloss et al., 2012), where the particles considered were soft spheres and the interactions between the particles in contact were taken into account. The normal repulsive forces were represented by a spring and dashpot in parallel. When two particles collided, the normal component of the velocity, $v_n$, between two particles was reduced to $-ev_n$, where $e$ was the coefficient of restitution set to a value of 0.1. We also considered the effect of friction on the change in the tangential component of the velocity between two particles based on Coulomb's friction law (the coefficient of friction is set to 0.5) and the gravitational force (the gravitational acceleration $g$ was set to 9.81 m/s$^2$), but cohesive force was not included in the simulation for simplicity.

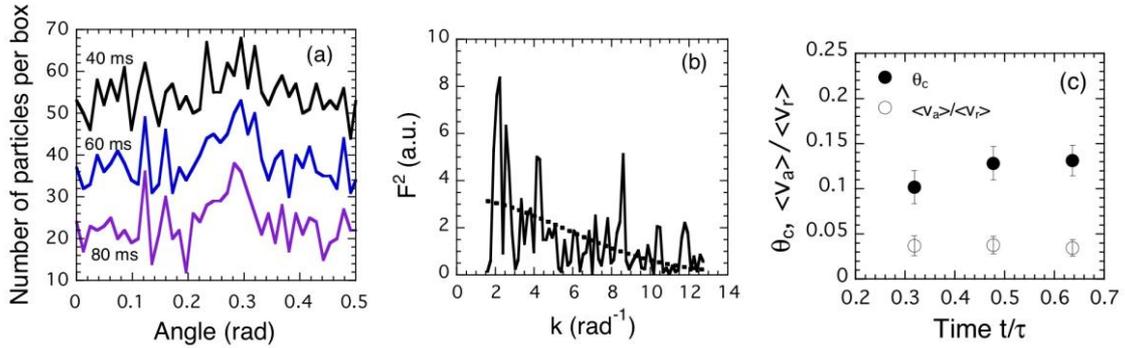

**Fig. 7.** The results of the numerical simulation. (a) Particle distributions as a function of azimuthal angle in the ejecta curtain with height between 8.3−12 cm at 40, 60, and 80 ms after impact. The profiles at 40 and 60 ms are offset to avoid crossover adding +30 and +10, respectively. (b) Power spectrum of the distribution at 60 ms. Fitting curve is also shown as a dotted curve. Spectra at 40 and 80 ms are shown in Fig. S4 in the SM. (c) Characteristic angle $\theta_c$ for the particle distributions as a function of time. The ratio $<v_a>/<v_r>$ is also plotted. The errors in $\theta_c$ and $<v_a>/<v_r>$ are from those in the fitting and in the averages over all particles, respectively.

The simulation considered $1.14 \times 10^6$ spherical particles with a radius of 0.7 mm as a target, whose density, Young modulus, and Poisson ratio were set to values of 2.7 g/cm$^3$, 9.4 MPa, and 0.17, respectively. Initially, these particles were accumulated at random in a hypothetical rectangular box with a size of 20 cm × 20 cm along the horizontal plane and a height of ~15 cm, and then gravitationally settled in a rectangular box with a size of 20 cm × 20 cm along the horizontal plane and a height of 7 cm. We considered a hypothetical sphere with a radius of 1.3 cm as projectile, in which $10^3$ spherical particles with a radius of 0.7 mm (whose density, Young modulus, and Poisson ratio were set to values of 1.0 g/cm$^3$, 9.4 MPa, and 0.17, respectively) were accumulated. This sphere collided vertically at the center of the target surface with a velocity of 100 m/s (see the SM for an animation). We considered the distribution of particles in the ejecta curtain between the heights of 8.3−12 cm from the bottom surface of the target box. The ejecta curtain was divided azimuthally into 512 boxes. A part of the distributions at 40, 60, and 80 ms after the impact is shown as a function of the azimuthal angle (Fig. 7a). We performed Fourier transformation, and the power spectrum was obtained as shown in Fig. 7b as a function of frequency $k$ in units of rad$^{-1}$ (the spectra at large $k > 12.8$ rad$^{-1}$ (wavenumbers > 80) and small $k < 1.5$ rad$^{-1}$ (wavenumbers of 1−9) are not considered) (Each power spectrum of the distributions shown in Fig. 7a at 40, 60 and 80 ms is shown in Fig. S4 in the SM). Though the spectrum fairly scatters, the curve appears to decrease with $k$. The characteristic angle $\theta_c$ was obtained by fitting a Gaussian function (a fitting curve is shown in Fig. 7b) and compared with the ratio $<v_a>/<v_r>$, where $<v_a>$ and $<v_r>$ are the absolute azimuthal and radial velocity components averaged over all particles, respectively. We considered the average of the absolute azimuthal component $<v_a>$ as the fluctuation velocity $\Delta v$. Figure 7c shows $\theta_c$ and $<v_a>/<v_r>$ as a function of time normalized by a characteristic crater formation time $\tau = (D_c/g)^{1/2}$, where crater (rim-to-rim) diameter $D_c$ was ~15.5 cm in this simulation. It appears that $\theta_c$ is almost constant with time at intermediary stages of crater formation; this is consistent with the experimental result. The ratio $<v_a>/<v_r>$ is also almost constant and ~1/3−1/4 of $\theta_c$. This confirms that $\theta_c$ and the ratio between the fluctuation and radial velocity components are in the same order of magnitude.

*4.3. Dependence of $\theta_c$ on impact conditions*

The experimental result shows no dependences of the angle $\theta_c$ on impact velocity. In cratering, impact causes a shock wave and, after the initial shock wave has dissipated, the excavation flow field is established (e.g., Melosh, 1985; 1989; Wada et al., 2006). Though the excavation flow field created by the shock wave's passage should be complex, our experimental result indicates that the ratio between the fluctuation and radial velocity components in the excavation flow field is quite similar regardless of impact velocity.

No difference in $\theta_c$ among the results of GB50, GB100_2, and GB300 reveals no dependence on particle sizes. Furthermore, the difference of $\theta_c$ between SS70 and GB50 is small within the scattering of the data (Fig. 5). Therefore, the effect of particle shape is also small. Since cohesive forces such as van der Waals and capillary forces between particles depend on size and shape of particles, these results imply that the cohesive forces and closely related elastic properties of particles do not influence $\theta_c$ representing the distance between clusters, i.e., the structure of patterns. However, this does not mean that inelastic collision between particles is not important. If the coefficient of restitution $e$ is not sufficiently low, the particles diffuse and the pattern changes temporary; i.e., $e$ plays key roles in the formation of distinct and persistent clusters. In fact, in freely falling granular streams in a vacuum (Möbius, 2006; Royer et al., 2009), Royer et al. (2009) show that cohesive forces are responsible for the occurrence of discrete, compact, and not transient clusters and, in experiments using spherical particles rolling on a smooth surface driven by a moving wall (Kudrolli et al., 1997), energy dissipation due to inelastic collisions plays a role in the formation of not dispersed clusters. Thus, we conclude that the elastic properties of particles determine whether distinct and persistent clusters are formed or not but have little influence on the structure of consequent patterns.

In this paper we used identical particles in size in each shot but the surface of natural bodies would not consist of identical particles in size but the mix of various sized particles. When target granular materials include larger particles, the fluctuation velocity of particles in the excavation flow would be different and also coalescence of particles should be promoted due to their larger collision cross-section. As a result, the patterns may be different. In fact, the recent experiments with the target consisting of two kinds of particles in size (0.1 mm and 1 or 4 mm) mixed at almost the same weight % show that large inclusions disturb excavation flow as obstacles and cause different flow patterns such as drags and spurts, but these patterns are temporary and not periodic (Kadono et al. 2019). On the other hand, when the amount of small particles is relatively large, the pattern would become similar to the one observed in our experiments. However, the systematic investigation of the effects of size distributions is insufficient and hence, should be done as a future work.

**5.Conclusion**

We investigated the patterns observed in ejecta curtain induced by hypervelocity impacts and the dependences of the patterns on the impact conditions such as impact velocity, particle size, and shape. We considered a quantity that characterizes the pattern, the angle $\theta_c$ defined as the ratio of the characteristic length of the pattern obtained by Fourier transformation to the distance from the impact point. This angle $\theta_c$ was found to be constant through time after the impact, suggesting that the pattern formation finished by the observation time. We evaluated an angle at which cluster growth stopped assuming that mutual collision of particles with fluctuation velocity resulted in clusters and showed that this angle and the ratio of the fluctuation velocity to the radial velocity were in the same order of magnitude. We confirmed this relation in the results of experiments and the numerical simulations. The experiments also show no dependence of $\theta_c$ on impact velocity. This indicates that the ratio between the fluctuation and radial velocity components in excavation flow does not depend on impact velocity. Moreover, the independences on particle size and particle shape suggest that the cohesive forces would not contribute the structure of the patterns.

The formation process of the pattern observed in the ejecta curtain can be summarized by the following: (1) after the initial shock wave dissipated and the excavation flow field was established, the particles acquired fluctuation velocity in the horizontal direction and accumulated through inelastic mutual collision, (2) clustering stopped when a characteristic angle of the pattern became at a certain value, which was the same in the order of magnitude as the ratio between the fluctuation and radial velocities, and (3) the pattern expanded geometrically and was observed in the ejecta curtain.

The patterns in ejecta curtain induced by hypervelocity impacts are consistent with those in the previous study at lower impact velocities (Kadono et al. 2015). Since the previous results are consistent with lunar crater-rays, we can conclude that the patterns observed in ejecta curtain induced by hypervelocity impacts are also consistent with the lunar crater-rays. Crater-rays are also observed on other airless bodies such as Mercury (e.g., MESSENGER website: http://messenger.jhuapl.edu/Explore/Science-Images-Database/By-Topic/topic-64.html) and asteroids (e.g., Williams et al. 2014). The comparison between the ray patterns on such bodies would verify the formation process that we proposed. For evaluation of such natural crater-ray systems on complex topography, however, the quantitative methods capable of analyzing the patterns in two-dimensions would be necessary to assess the similarity and difference between the patterns of ejecta deposit more precisely than our analysis in one-dimensional.

**Acknowledgments**

We thank M. Arakawa for the discussion based on his unpublished data and M. Koga for supporting data analysis and creating some figures. The authors are also grateful to two anonymous reviewers for helpful comments. This work was supported by ISAS/JAXA as a collaborative program with the Hypervelocity Impact Facility.